\documentstyle[11pt,COIMBRA_pasp,epsf]{article}      

\begin{document}

\title{The BaSeL library as a basic tool to provide fundamental 
stellar parameters of future space missions: COROT and GAIA.}

\author{E. Lastennet}
\affil{Depto. de Astronomia, UFRJ, Rio de Janeiro, Brazil}

\author{T. Lejeune}
\affil{Observat\'orio Astron\'omico da Universidade de Coimbra, Portugal}

\author{F. Cuisinier}
\affil{Depto. de Astronomia, UFRJ, Rio de Janeiro, Brazil}

\begin{abstract}
COROT and GAIA are two future major space missions directly connected 
to most of the stellar astrophysic questions, from stellar physics to
evolution of galaxies. 
We describe a project for the preparation of these two missions 
by using the possibilities of the {\em ``BaSeL'' models}, a library
of theoretical stellar energy  distributions, to provide
automatically the fundamental stellar parameters of the candidate stars. 
We present the results already obtained for the stars of the COROT main
programme. 

\end{abstract}

% Keywords should be included, but they are not printed in the hardcopy.

\keywords{fundamental parameters, binary systems}

\section{Brief description of the BaSeL models}
%%%%%%%%%%%%%%%%%%%%

The Basel Stellar Library (BaSeL) is a library of theoretical spectra 
corrected to provide synthetic colours consistent with empirical colour-temperature 
calibrations at all wavelengths from the near-UV to the far-IR (see Cuisinier 
et al. 1996 for the correction procedure, and Lejeune et al. 1998 and references 
therein for a complete description). 
These model spectra cover a large range of fundamental parameters 
(2000 $\leq$ T$_{\rm eff}$ $\leq$ 50,000 K, $-$5 $\leq$ [Fe/H] $\leq$ 1 
and $-$1.02 $\leq$ log g $\leq$ 5.5) and hence 
allow to investigate a very large panel of multi-wavelength astrophysical questions. 
\\
The BaSeL library spectra have been calibrated directly for standard dwarf and 
giant sequences at solar abundances and using UBVRIJHKLM broad-band photometry, 
and are hence expected to provide excellent results in these photometric bands.
Since they are based on synthetic spectra, they can in principle be used in many 
other photometric systems taken either individually or simultaneously, and this 
is another major advantage of these models. 
Moreover, their photometric calibrations are regularly updated and extended 
to an even larger set of parameters (see Lejeune et al. 1998 and references 
therein). In the two next sections we describe how we intend to apply this 
library to COROT and GAIA target stars.

%%%%%%%%%%%%%%%%%%%%
\section{COROT: major asteroseismology space mission}
%%%%%%%%%%%%%%%%%%%%

COROT ({\bf CO}nvection and {\bf ROT}ation) is a space experiment dedicated to 
ultra high precision, wide field, relative stellar photometry, for very long 
continuous observing runs on the same field of view.
It has two main scientific programs working simultaneously on adjacent regions 
of the sky: asteroseismology and search for extrasolar planets. 
To perform an optimal selection among the potential targets 
it is necessary to know their fundamental parameters. 
One of the objectives of the preliminary study of target characterization for the 
central seismology programme of COROT (Ligni\`eres et al. 1999) 
was to choose the appropriate method(s) to determine the basic parameters of 
the potential targets of this programme (T$_{\rm eff}$, log g, abundances, vsini,
multiplicity). One of these methods is based on the BaSeL library. 

Given a set of effective temperature, surface gravity and metallicity 
(T$_{\rm eff}$, log g, [Fe/H]), the BaSeL models provide various colours that
can  be directly compared with stellar populations.  
The inverse method (Lastennet et al. 1999) is as well very useful to derive 
the atmospheric parameters from the observed colours: Lastennet et al. (1999) 
applied this method to a sample with photometry in the Str\"omgren
system, and obtained very good results from the BaSeL models, in agreement 
with accurate HIPPARCOS data. 

The capability of the BaSeL models to derive simultaneously or
individually  the (T$_{\rm eff}$, log g, [Fe/H]) parameters for the main 
programme stars of the COROT mission has already been shown in Lastennet 
et al. (2001). 
For illustration purpose, we show some of the final results 
we obtain in Figure 1. 
We propose to develop in the near future an automatic tool
based on the method of  Lastennet et al. (1999) to complete the facilities of
the new "BaSeL interactive  server", the web version of the BaSeL models
hosted by the Coimbra Observatory since the end of 2000 ({\tt
http://www.astro.mat.uc.pt/BaSeL/}, see also Lejeune \& Schaerer, 2001
for  details). This method would be applied to the $\sim$ 1000 remaining 
potential targets of the COROT exploratory programme. 

%%%%%%%%%%%%%%%%%%%%
\section{GAIA: ESA Cornerstone space mission}
%%%%%%%%%%%%%%%%%%%%

The GAIA mission, an ESA (European Space Agency) cornerstone mission, 
has been designed to solve many of the
most difficult and deeply fundamental challenges in modern
astronomy: to determine the composition, formation and
evolution of our Galaxy.
GAIA will 
provide unprecedented positional and radial velocity measurements with 
the accuracies required to produce a stereoscopic and kinematic census of 
about one billion stars in our Galaxy (this amounts to about 1 \%  
of the Galactic stellar population) and throughout the Local Group.  
Combined with astrophysical information for each star, provided by on-board 
multi-colour photometry, these data will have the precision necessary to
quantify the early formation,
and subsequent dynamical, chemical and star formation evolution of
the Milky Way Galaxy.
Additional scientific products include detection of new binary systems,
brown dwarves, extragalactic objects (more than 1 million
galaxies, 5.10$^6$ quasars, 10$^5$ supernovae etc...), with
crucial implications for stellar and galactic physic,
galactic structure, distance scales in the Local Group, etc... \\
The core science case for GAIA requires measurement of luminosity,
T$_{\rm eff}$, mass, age, and composition for the stellar populations in our
own Galaxy and in its nearest galaxy neighbours. These quantities
can be derived from the spectral energy distribution of the stars, through
multi-band photometry. \\
Nonetheless, while many photometric systems already exist (e.g. Johnson-Cousins,
Geneva, Str\"omgren, RGU, Washington, etc...) none satisfy all the GAIA
requirements. The GAIA photometric system must be able to classify stars
across the entire Hertzsprung-Russell diagram, as well as to identify peculiar
objects. It must be able, for example, to determine
temperatures and reddening for OBAFG stars (needed as tracers of Galactic
spiral arms and as reddening probes), temperatures and abundances
for late-type giants and dwarfs, abundance of Fe and $\alpha$-elements,
etc....
Thus it is necessary to observe a large spectral domain, extending from
the UV to the far-infrared.  \\
Provided that the new pass-bands of the GAIA space mission
(e.g. the 4 broad bands and 11 intermediate bands covering the spectral range
280 to 920 nm, see Cayrel et al. 1999 and
{\tt http://wwwhip.obspm.fr/gaia/photometrie.\-html})
are implemented in the BaSeL models, the new tool described in the
previous section will provide automatically
(T$_{\rm eff}$, log g, [Fe/H]) estimates for the stars observed by GAIA.

\section{Concluding remarks}
COROT and GAIA are two next generation space missions which
should provide essential results for stellar evolution theory.
Here we present a proposition to develop for these two missions an automatical
method, already used with success for COROT potential targets (Lastennet et
al. 2001), for a systematic determination of fundamental parameters from
BaSeL synthetic multi-photometry.
This new tool should be publicly available on the following  web site
{\tt http://www.astro.mat.uc.pt/BaSeL/} by the end of 2001.

\acknowledgments
EL acknowledges support from CNPq and FAPERJ.

%%%%%%%%%%%%%%%%%%%%
\begin{figure}
%\plotfiddle{corot_basel.ps}{6.cm}{-90}{50}{65}{-230}{200}
\plotfiddle{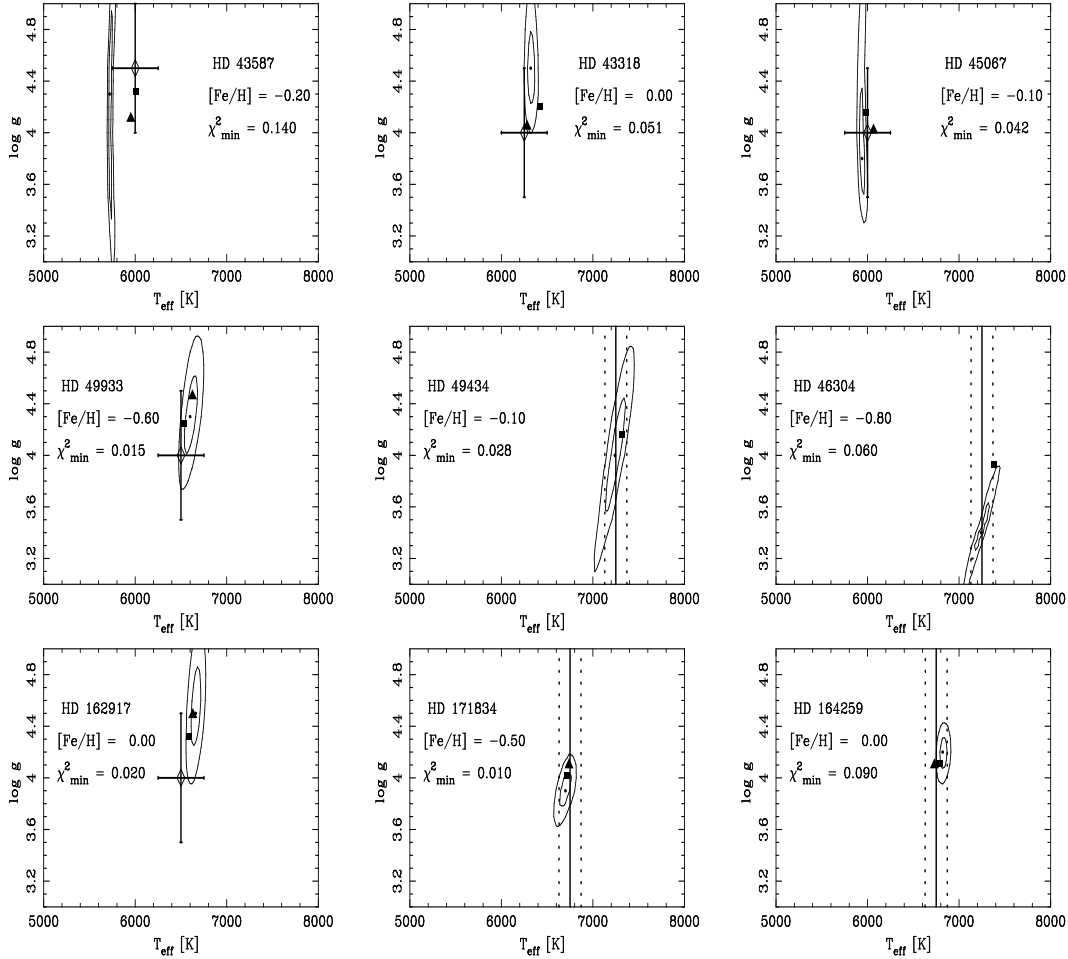}{11.cm}{-90}{50}{65}{-230}{380}
 \caption{
 \small
Results in the (T$_{\rm eff}$, log g) diagram for
9 potential targets of the COROT central seismology programme.
For each star, the 1- and 2-$\sigma$ contours are displayed in a [Fe/H]
$=$ constant plane,  corresponding to the best simultaneous (T$_{\rm eff}$,
[Fe/H], log g) solutions derived  from the BaSeL models in order to fit
simultaneously the observed  photometric values (B$-$V),
(U$-$B) and (b$-$y). An estimation of the quality of the best fit
($\chi^2$-value) is also quoted in each panel.  The results projected in the
T$_{\rm eff}$-log g planes from the spectroscopic analysis ({\it diamond with
error bars}, or {\it solid plus two dotted lines})  as well as from the
"Templogg" programme ({\it square}),  and Marsakov \& Shevelev (1995) ({\it
triangle}) are also shown for comparison.
\normalsize
 }
\label{f:corot_basel}
\end{figure}
%%%%%%%%%%%%%%%%%%%%

\end{document}